\documentclass[aps, prb, showpacs, twocolumn]{revtex4}
\usepackage{hyperref}
\usepackage{graphicx}
\usepackage{amsmath}
\usepackage{amssymb}
\usepackage{amsfonts}   
\usepackage{multirow}
\usepackage{color}

\begin{document}
\title{Effect of dilute, strongly pinning impurities on  charge density waves}

\author{Jun-ichi Okamoto}
\email{okamoto@phys.columbia.edu}
\affiliation{Department of Physics, Columbia University, 538 West 120th Street, New York, New York 10027, USA}
\altaffiliation{Present address: Institut f\"{u}r Laserphysik, Universit\"{a}t Hamburg, Luruper Chaussee 149, D-22761 Hamburg, Germany}

\author{Andrew J. Millis}
\affiliation{Department of Physics, Columbia University, 538 West 120th Street, New York, New York 10027, USA}

\date{\today}

\begin{abstract}
We study theoretically the effects of strong pinning centers on a charge density wave in the limit that the charge density wave coherence length is shorter than the average inter-impurity distance. An analysis based on a Ginzburg-Landau model shows that long range forces arising from the elastic response of the charge density wave induce a kind of collective pinning which suppresses impurity-induced phase fluctuations leading to a long ranged ordered ground state. The effective correlations among impurities are characterized by a length scale parametrically longer than the average inter-impurity distance. The thermal excitations are found to be gapped implying the stability of the ground state. We also present Monte Carlo simulations that confirm the basic features of the analytical results. 

\end{abstract}

\maketitle
\section{Introduction}

Recent scanning tunnelling microscopy (STM) data \cite{Okamoto2014} suggests that the charge density wave (CDW) in NbSe$_2$ is a topologically ordered Bragg glass at least on the length scales ($\sim 30$ unit cells) accessible in the experiment. Such a glass phase with a power law density correlation was predicted  theoretically  \cite{Feigel'man1989, Nattermann1990, Bouchaud1991, Bouchaud1992, Korshunov1993, Giamarchi1994, Giamarchi1995, Giamarchi1997, Rosso2004} for three dimensional systems with a high density of weak pinning centers. However, analysis of the STM data suggests  that the impurities in NbSe$_2$ are strong pinning centers.\cite{Okamoto2014}  In this situation, conventional theories \cite{Efetov1977, Fukuyama1978, Lee1979} predict  short-range order with a correlation length $\xi$ of the order of the average inter-impurity distance $l$ and with  a proliferation of topological defects,\cite{Giamarchi1995,Giamarchi1997} in apparent contradiction to the experimental result. The NbSe$_2$ sample studied in Ref.~\onlinecite{Okamoto2014} was found to have  $\xi \approx 1$ nm and $l \approx 5$ nm, so the pinning centers are dilute on the scale set by the CDW coherence length. This dilute-pinning limit has not been extensively studied in the literature. Here, we present a theoretical reexamination of the dilute impurity limit which indicates that in this limit unconventional and apparently previously overlooked physics occurs. This paper amplifies and extends results presented in Ref.~\onlinecite{Okamoto2014} in the context of an analysis of experimental data. 

Our work builds on ideas introduced many years ago by Abe,\cite{Abe1985, Abe1986} who  pointed out the importance of  the ratio between the coherence length $\xi$ and the average inter-impurity distance $l$. A charge density wave is characterized by an amplitude and a phase. A strongly pinning impurity constrains the charge density wave phase to take a particular value on the impurity site. Abe presented a scaling argument indicating that for impurities acting as strong pinning centers one should distinguish two cases. When $\xi > l$ the CDW phase simply interpolates between  the values preferred by impurities leading to a correlation length of order  $l$. On the other hand, when $\xi < l$, Abe found that the preferred solution is for the phase to be very slowly varying, except in small regions of size $\xi$ around each impurity where the phase varies rapidly to accommodate the values preferred by the impurities. Abe's analysis demonstrates the importance of the $l\gg\xi$ limit, but as will be seen below his analysis of this case is not completely correct. Here, we reconsider the $l\gg\xi$ limit, finding that a screening effect arising from the long ranged nature of elastic forces makes the net effect of impurities weak at long length scales. The resulting state is stable, in the sense of having an energy gap for fluctuations.  We also present numerical results confirming the main aspects of our arguments.

The rest of this paper is organised as follows: Sec.~\ref{Model} presents the model we study and a formal solution. Sec.~\ref{RandomPhase} presents the details of an analytical solution of a simplified model in which the impurities are placed on the sites of a regular lattice, with randomness only entering in the values of the phases preferred by impurities; effects of vortices and the periodicity of the pinning potential are also neglected.  Thermal excitations are discussed in Sec.~\ref{Excited}. Sec.~\ref{Numerics} presents a numerical analysis of the ground state of the  model defined in Sec.~\ref{Model}, which confirms that the predictions of the simplified model apply to the full model. Sec.~\ref{MC} presents Monte Carlo analyses that confirm the main aspects of the analytical results and Sec.~\ref{Conclusions} is a summary and conclusion. 

\section{Model and Formal Solution \label{Model}}

Although our work is motivated by  experiments on  NbSe$_2$, \cite{Okamoto2014} which has a three-component order parameter,  for theoretical simplicity we consider here  a generic one-component CDW in which the density modulation $\delta\rho(\vec{x})=\Re\left[ A(x)e^{i\vec{Q}\cdot\vec{x}+\phi(\vec{x})}\right]$, and we take the amplitude $A$ to be constant. We also assume that all impurities have the same pinning strength $V$, with randomness entering via the value $\theta_a$ of the phase preferred by the impurity at position $\vec{x}_a$. We believe that the qualitative behaviour we uncover is relevant also to the more complicated three-component, amplitude-varying case of NbSe$_2$.

Within our assumptions, the Ginzburg-Landau (GL) energy density is reduced to the phase-only model,\cite{Fukuyama1978} 
\begin{equation}
E(\vec{x}) = \frac{1}{2} \rho_S\left[ \vec{\nabla } \phi(\vec{x}) \right]^2 - V \sum_{a}\cos \left[ \theta_a - \phi (\vec{x}) \right]\delta\left(\vec{x}-\vec{x}_a\right),
\label{phase_model}
\end{equation}
where $\rho_S$ is the CDW phase stiffness (elastic constant). We will focus on the three dimensional case, since both experimental \cite{Moncton1977} and theoretical \cite{Calandra2009} studies indicate that the CDW in NbSe$_2$ is not extremely anisotropic.  

If Eq.~\eqref{phase_model} is taken literally as a model defined in the continuum, the phase variable is non-compact and vortices or antivortices are forbidden. However the model applies only at length scales longer than the CDW coherence length $\xi$ and physics relevant on the length scale of $\xi$ (for example amplitude fluctuations) allows vortex-antivortex excitations and makes the phase field compact.    The model defined by Eq.~\eqref{phase_model} arises also as the long wavelength limit of the lattice  XY model of spins on a lattice with dilute random magnetic fields, but the short wavelength physics of this model may differ in details from that of Eq.~\eqref{phase_model}.

Minimization of Eq.~\eqref{phase_model} shows that $\phi$ obeys the Laplace equation, $\nabla^2 \phi  = 0$ for all $\vec{x}$ not within a distance $\xi$ of an impurity site.  Under these assumptions, the most general solution of the Laplace equation  is 
\begin{equation}
\phi (\vec{x}) = \sum_a \frac{\bar{\theta}_a \xi}{\left|\vec{x} - \vec{x}_a \right|},
\label{ansatz}
\end{equation}
where the  $\bar{\theta}_a$ are parameters and Eq.~\eqref{ansatz} applies only if $\left|\vec{x} - \vec{x}_a \right|>\xi$ for all $a$.  For $\left|\vec{x} - \vec{x}_a \right|<\xi$ we regularize the formally divergent term in the sum as  $\bar{\theta}_a$. To determine the parameters $\bar{\theta}_a$ we substitute Eq.~\eqref{ansatz} into Eq.~\eqref{phase_model} and minimize the result with respect to $\bar{\theta}_a$. We focus here on  the strong pinning limit, in which  we expect that the phase $\phi(x\rightarrow x_a)\approx \theta_a$, so we can expand the impurity potential up to second order. We find
\begin{equation}
\frac{E}{V} \simeq \frac{\epsilon}{2}\sum_{ab} K_{ab} \bar{\theta}_a  \bar{\theta}_b + \frac{1}{2} \left(\theta_a  - \sum_b K_{ab} \bar{\theta}_b +2\pi n_a \right)^2 
\label{Eofthetabar}
\end{equation}
with 
\begin{equation}
K_{ab} \equiv  \delta_{ab} +\left(1-\delta_{ab}\right)\xi / |\vec{x}_a - \vec{x}_b|
\label{Kdef}
\end{equation}
and $\epsilon = 4\pi \rho_S \xi/V \ll 1$. The integer $n_a$ accounts for the periodicity of the cosine potential. For a single impurity, $n_a \neq 0$ simply increases the elastic energy compared to $n_a = 0$. We expect that with many impurities, $n_a \neq 0$ solutions are generically  energetically expensive, so  we will focus on $n_a = 0$ solutions for our initial analysis, and return to the effects of nonzero $n_a$ below (See Sec. \ref{Numerics}). 

Eq.~\eqref{Eofthetabar} is an analogue of a Coulomb gas with the constraint that the potential take specific values on particular sites. Minimizing Eq.~\eqref{Eofthetabar}  in terms of $\bar{\theta}_a$, we find
\begin{equation}
\bar{\theta}_a =  \sum_b J_{ab} \theta_b ,
\label{thetabar}
\end{equation}
where
\begin{equation}
J=\sum_b \left(\epsilon I + K \right)^{-1}_{ab} \theta_b 
\label{Jdef}
\end{equation}
and $I$ is the identity matrix. Substituting Eq.~\eqref{thetabar} into Eq.~\eqref{Eofthetabar} and setting $n_a=0$ yields a total energy $E = 2\pi \rho_S \xi \sum_{ab} J_{ab}\theta_a  \theta_b$. The phase configurations given by the above expressions are  in general found to be smooth although rare vortex-antivortex pairs  occur. These are allowed because the regularization of $1/|x-y|$ introduced above in effect allows vortex cores of size $\xi$. As will be seen below, vortices are in practice  induced by particular impurities that favor phase values very different from the average. In the calculations we have performed, however, the vortices are very dilute (typically $\sim 0.2\%$; see below).

\section{Simplified Model \label{RandomPhase}}

The key physics of Eqs.~\eqref{ansatz} and \eqref{Eofthetabar} is that because of the long-range nature of the elastic forces in a  Laplacian problem, the phase at a given site is determined by the collective response at many impurity sites. To gain analytical insights into this physics we consider a simplified model in which the the impurities on the sites of  a cubic lattice with  lattice constant $l$ and the randomness enters only through the values of the parameters $\theta_a$.  The kernels $K$ [Eq.~\eqref{Kdef}] and $J$ [Eq.~\eqref{Jdef}] are then defined on the points of the lattice of impurities and taking $\vec{p}$ to be a vector within the Brillouin zone of this lattice ($|p_{x,y}|<\pi/l$), we find
\begin{equation}
K (p)=\frac{1}{r_\text{TF}^2p^2}+1 \cdots,
\label{kernelmomentum}
\end{equation}
while 
$J$ has the familiar screening form:
\begin{equation}
J (p) \simeq \frac{p^2 r_\text{TF}^2 }{p^2 r_\text{TF}^2 + 1}+\cdots
\label{J}
\end{equation}
Here $r_\text{TF}=\sqrt{l^3 /4\pi \xi}$ is a characteristic length that is parametrically larger than $l$, and $\cdots$ denotes terms that are smaller by powers of $\xi/l$ and of $\mathcal{O} (\epsilon)$ relative to the terms that have been retained.  Abe \cite{Abe1986} considered a similar scenario but assumed a simple exponential form of $J(x)$ instead of the screening form given by Eq.~\eqref{J}.

\begin{figure}[!tb]
  \begin{center}
   \includegraphics[width = 0.8\columnwidth ]{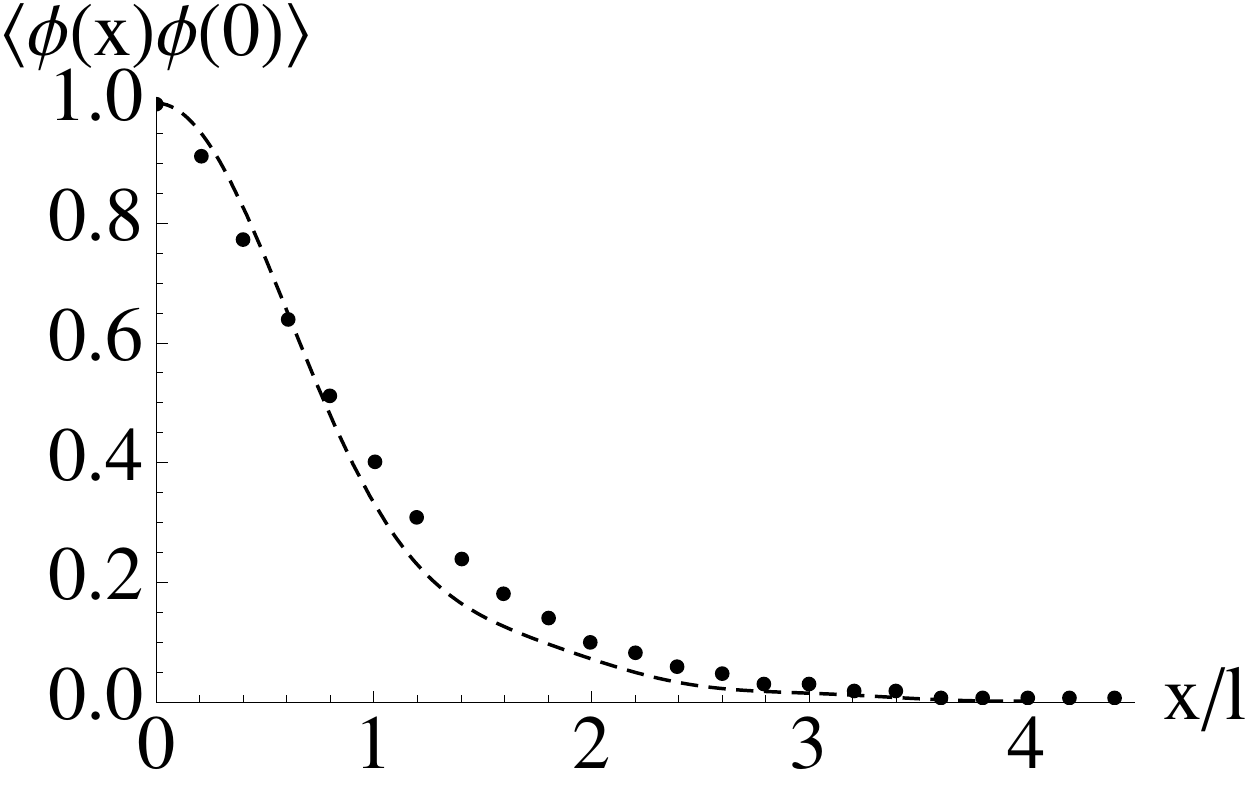}
   \caption{Points: normalized phase-phase correlation obtained by averaging numerical solutions of Eq.~\eqref{thetabar} over 50 realizations of randomly placed impurities with  $l =5.0 \xi$ and linear sample size $L=60$. Dashed line: normalized phase-phase correlation calculated from Eq.~\eqref{phase-corr}.}
    \label{phi_auto}
  \end{center}
\end{figure}

\begin{figure*}[!t]
  \begin{center}
   \includegraphics[width = 1.8\columnwidth]{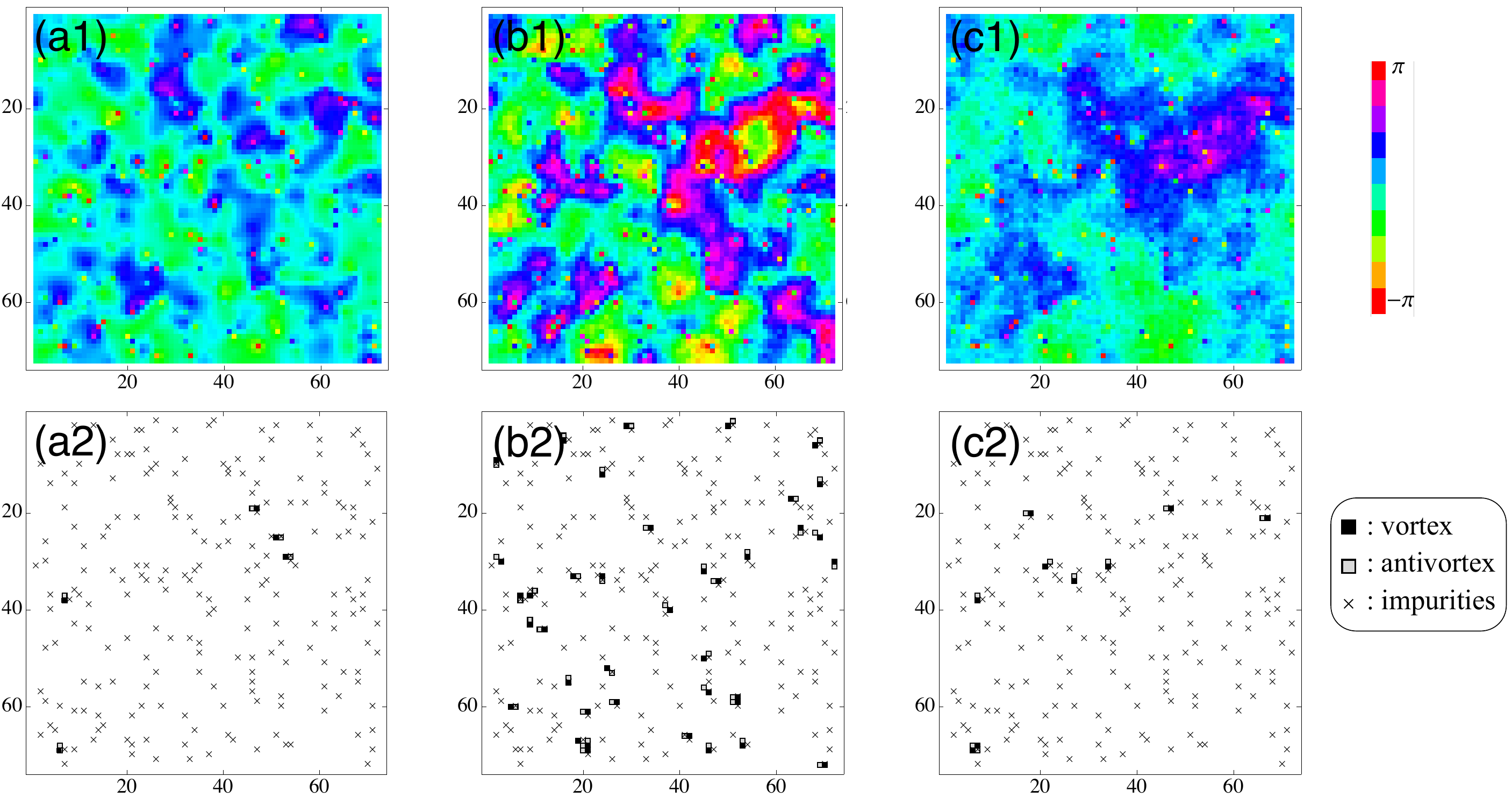}
   \caption{(Color online) Upper panels are typical phase configurations for the same impurity distributions obtained by three methods: (a) Eqs.~\eqref{kernelmomentum} and \eqref{J} without $n_a$ variations, (b) the same as (a) but with $n_a$ variations, and (c) Monte Carlo simulations. The vortex densities $n_V$ are $0.2\%$, $1.4\%$, and $0.5\%$ respectively. Lower panels correspond to the vortex configurations of the same field of views. }
    \label{phase}
  \end{center}
\end{figure*}

The physics encoded in Eqs.~\eqref{kernelmomentum} and \eqref{J} is that the physical CDW phase at any given point is determined collectively, with important contributions from impurities very far away, and this collective nature of the pinning suppresses the impurity-induced phase fluctuations. Mathematically, the phase parameters $\bar{\theta}_a$ that determine the effect on the CDW phase of an  impurity on site $a$ are not independent, and indeed have long ranged (Coulombic) correlations even for delta-correlated randomness $\langle \theta_a\theta_b \rangle \sim \delta_{ab}$.  Explicitly, use of Eqs.~\eqref{thetabar} and \eqref{J} leads to (we have approximated the Brillouin zone of the impurity lattice as a sphere of appropriate radius, because we are interested in the long length scale behaviour)
\begin{equation}\label{phase-corr}
\begin{split}
\langle \phi(\vec{x})\phi(\vec{y}) \rangle &= \sum_{ab}\frac{\left\langle \bar{\theta}_a\bar{\theta}_b\right\rangle}{|\vec{x}-\vec{x}_a||\vec{y}-\vec{x}_b|} \\
&\approx  \frac{8\xi^2 \pi^2}{3l^3}  \int _0^{2\pi/l}dp  \frac{p   \sin \left( p \left|\vec{x}-\vec{y}\right| \right)}{\left|\vec{x}-\vec{y}\right| \left( p^2+ r_\text{TF}^{-2} \right)^2},
\end{split}
\end{equation}
indicating that the phase-phase correlation is characterised by $r_\text{TF}$ and, crucially, does not diverge as $x\rightarrow y$ indicating that the root mean square phase fluctuation induced by the impurities remains bounded. Thus, the average of the  CDW order parameter  $\langle e^{i \phi}\rangle = e^{-\frac{1}{2} \langle \phi^2 \rangle}$ is non-zero, so that even in the presence of impurities the model has long-range order in the ground state.  In other words, the correlation function defined as
\begin{equation}
C(\vec{x}) \equiv \langle \vec{s}(\vec{x}) \cdot \vec{s}(0)\rangle = \langle e^{i \phi(\vec{x})} e^{-i \phi(0)}\rangle
\label{Cx}
\end{equation}
with $\vec{s} = (\cos \phi, \sin \phi )$ remaining non-zero as $x\rightarrow \infty$, since $C(x \rightarrow \infty)\rightarrow | \langle e^{i\phi} \rangle|^2$.

\section{Excited States and Thermodynamics \label{Excited}}

We now consider states that are not the ground state. These may be generated in two ways. One may consider states of the form of Eq.~\eqref{ansatz} but with parameters $\bar{\theta}_a$ that do not satisfy Eq.~\eqref{thetabar}. These solutions will have very high energy, because they violate the pinning conditions. Alternatively, one may consider solutions $\phi(x)$ which correspond to elastic excitations about the ground state but with the pinning condition satisfied. To obtain such solutions we write
\begin{equation}
\phi (\vec{x}) =  \Delta (\vec{x}) + \sum_a \frac{\bar{\theta}^\prime_a \xi}{\left|\vec{x} - \vec{x}_a \right|}.
\label{ansatz2}
\end{equation}
$\Delta$ does not obey the Laplace equation, because the configuration does not minimize the energy.  The $\bar{\theta}^\prime_a$ [whose dependence on $\Delta(x)$ is not explicitly notated here] are determined by minimizing the energy for fixed $\Delta(x)$, in particular insuring that  the impurity pinning condition is fulfilled. Substituting Eq.~\eqref{ansatz2} into Eq.~\eqref{phase_model} we obtain
\begin{multline}
E[\Delta(x)] \simeq  \frac{\rho_S}{2} \int d^3 x \left( \nabla \Delta \right)^2 + \frac{V}{2}\sum_a (\Theta_a - \sum_b K_{ab} \bar{\theta}^\prime_b )^2\\
+ 2\pi \rho_S \xi \left[  \sum_{ab} K_{ab} \bar{\theta}_a^{\prime} \bar{\theta}^\prime_b + 2 \sum_{a} (\Delta_a -\Delta_0) \bar{\theta}^\prime_a \right] ,
\label{Eofdelta}
\end{multline}
where $\Delta_a=\Delta(x=x_a)$ and  $\Theta_a$ is the sawtooth function of $(\theta_a - \Delta_a)$ introduced to recover the periodicity $2\pi$, and taken to be in $[-\pi, \pi ]$. $\Delta_0$ is the  value of $\Delta$ at infinity. Minimizing Eq.~\eqref{Eofdelta} with respect to the  $\bar{\theta}^\prime_a$ gives,
\begin{equation}
\bar{\theta}^\prime_a = \sum_{b} J_{ab} \Theta_{b},
\end{equation}
and substituting this into Eq.~\eqref{Eofdelta} leads to (noting that terms of order $\epsilon$ and the multi branch structure of the sawtooth function are not important here)
\begin{equation}
E[\Delta(x)] = \frac{\rho_S}{2}\left[ \int  d^3 x \left( \nabla \Delta \right)^2 -4\pi\xi\sum_{ab} J_{ab} \Delta_a\Delta_b\right] + \text{const.}
\label{EofDelta}
\end{equation}
To analyze Eq.~\eqref{EofDelta}, we again consider the model in which the impurities are on a regular lattice of lattice constant $l$ and use Eq.~\eqref{J} for the Fourier transform of $J_{ab}$, obtaining finally
\begin{equation}
E =  \frac{\rho_S}{2} \int\frac{d^3 p}{(2\pi)^3} \frac{r_\text{TF}^2p^4}{r_\text{TF}^2p^2+1}|\Delta(p)| ^2.
\label{Eofdeltafinal}
\end{equation}
Since this is a positive definite, the ground state must have $\Delta (p \neq 0) =0$; long-range order is preserved. The thermal fluctuation is
\begin{equation}
\delta \phi \equiv \phi - \sum_a \frac{\bar{\theta}_a \xi}{\left|\vec{x} - \vec{x}_a \right|}, 
\end{equation}
where the second part is the ground state configuration with $\bar{\theta}_a$ given by Eq.~\eqref{thetabar}. The phase-phase correlation of the fluctuating part $\delta \phi$ is found to be 
\begin{equation}
\langle \delta \phi (\vec{x}) \delta \phi (0)\rangle \sim \frac{T}{\rho_S} \int \frac{d \vec{p}}{(2\pi)^3} e^{i \vec{p} \cdot \vec{x}} \frac{r_\text{TF}^2}{p^2 r_\text{TF}^2 +1} .
\end{equation}
This function decays exponentially in terms of $\vec{x}$, and, thus, the long-range order still survives. The finite energy to excite the fluctuation originates from the fact that impurities fix the fluctuations at impurity sites to zero, $\delta \phi (x_a) = 0$.

\section{Numerical Energy Minimization\label{Numerics}}
The analysis presented in section~\ref{RandomPhase} relied on three simplifications: placing the impurities on a regular lattice, so that the randomness only enters via the value of the phase preferred by the impurities, neglecting vortex configurations of the phase field, and ignoring the periodicity of the pinning potential under $\phi\rightarrow \phi+2\pi$.  In this section we  numerically investigate the consequences of relaxing these assumptions. We find that the simplifications do not  affect the qualitative physics. However, there is a notable quantitative difference between the full model and the approximate model of section~\ref{RandomPhase} arising from the effects of vortex-antivortex pairs, with correlation lengths changing by as much as a factor of two.  

The probability that vortex-antivortex pairs  will be produced depends both on the CDW stiffness $\rho_S$ and on the  details of the short distance physics, which control the core energy of a vortex. Specifically, the presence of a  vortex-antivortex pair near an impurity site can allow the phase to relax rapidly from the value preferred by the  impurity towards a background value, thus reducing the elastic energy. To estimate this energy we assume that the presence of the vortex-antivortex pair completely screens the impurity, so it is equivalent to removing one defect from the elastic energy. To estimate this effect  we use the screened Coulombic form of $K^{-1}$ and note that the $\theta_a$ are random variables. We find the elastic energy gain is roughly
\begin{equation}
E_\text{elastic} \simeq 2\pi\rho_S\xi (1+\epsilon)^{-1} \theta_a^2 + \mathcal{O} (\xi /l).
\label{E_vortex}
\end{equation}
The elastic energy gain is thus seen to depend on a property of the impurity, namely the phase change it produces, and on two properties of the CDW: the phase stiffness, and a short distance cutoff (here specified by the correlation length $\xi$. The net energy cost of creating a vortex-antivortex pair is thus the sum of $E_\text{elastic}$ and the core energy $E_\text{core}$ which general Ginsburg-Landau considerations suggest should also be of the order of $\rho_S\xi$.  Whether the total energy is positive or negative thus depends on $\theta_a^2$ and on an intrinsic property of the CDW, namely the ratio,
\begin{equation}
\kappa = 2\pi \rho_S \xi /E_\text{core}.
\label{kappa}
\end{equation}
$E_\text{core}$ depends on microscopic details beyond the scope of this paper. For example, in the XY model there is a significant elastic contribution\cite{Kosterlitz1973}  suggesting that  $E_\text{core}$ of the XY model  may be larger than the one of the phase-only model, effectively leading to a smaller value of $\kappa$; it is harder to make vortex pairs in the XY model.

\begin{figure}[!tb]
  \begin{center}
   \includegraphics[width = 0.9\columnwidth]{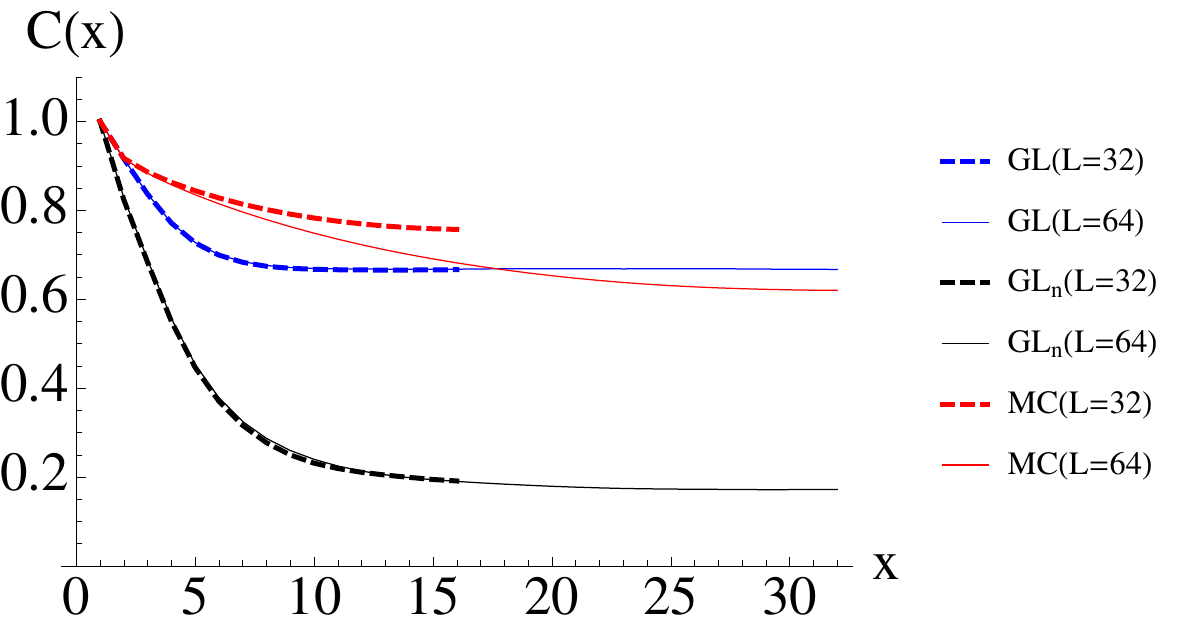}
   \caption{(Color online) Autocorrelations based on phase configurations obtained from three methods. Blue lines are based on the numerical solutions of Eqs.~\eqref{kernelmomentum} and \eqref{J} without $n_a$ variations (labelled as GL). The ones with $n_a$ variations are labeled as GL$_\text{n}$ and plotted by black lines. Red lines are from Monte Carlo simulations. Dashed (solid) lines corresponds to $L=32 \ (64)$.}
    \label{auto3}
  \end{center}
\end{figure}

Now we turn to the numerical energy minimization. We solved  the matrix equation \eqref{thetabar} without any simplifying assumptions by placing impurities at random on the nodes of a $L\times L\times L$ lattice of lattice constant $\xi$ with randomly chosen preferred phases. Note that putting the problem on a lattice is a particular way of imposing an ultraviolet cutoff and in particular permits vortices whose cores reside inside an elementary placquette.  We calculated $K$ and $J$ numerically but neglected the periodicity of the pinning potential by setting (at this stage) the $n_a=0$.  Fig.~\ref{phi_auto} compares the analytical expression for the phase-phase correlation, Eq.~\eqref{phase-corr}, to results obtained by averaging the numerical solution of  Eq.~\eqref{thetabar} over 50 realizations of the impurity potential. We see that the analytical expression agrees well with the numerical results. A typical phase configuration of $n_a = 0$ computed for an impurity density $n_\text{imp} = 4\%$ is shown in panel (a1) of Fig.~\ref{phase}. It exhibits a smooth phase modulation with  fluctuations of size $l$. The lower panel shows the positions of the impurities as well as the locations of the vortices and antivortices. We see that vortices and antivortices occur only in tightly bound pairs, always near impurity sites. The density of vortices is very low; $n_V = 0.2\%$.   Fig.~\ref{auto3} shows that the autocorrelation saturates at a non-zero value as the distance goes to infinity, with the value independent of the size of the system. We conclude that in this approximation model has long ranged order. 


We next consider the consequences of the periodicity of the pinning potential, by allowing configurations with  $n_a \neq 0$. After minimization over the $\bar{\theta}$ parameters the energy expression below Eq.~\eqref{Jdef} then becomes
\begin{equation}
E = 2\pi \rho_S \xi \sum_{ab} J_{ab} (\theta_a + 2\pi n_a ) (\theta_b + 2\pi n_b),
\label{na_energy}
\end{equation}
which we now minimize  as a function of $n_a$. To do this we search for the lowest energy state in the $\{ n_a \}$ space by a gradient annealing method. We start from the original state with $n_a = 0$, cycle through all $a$. For each $a$  we propose the  change $n_a \rightarrow n_a \pm 1$. If the change gives a lower energy, we accept it, and move onto the next $a$; otherwise, we do not update $n_a$ but still move on to the next $a$.  We repeat the procedure, cycling repeatedly over all impurity sites, until we get a converged result. For $4\%$ impurities, about 5$\%$ to $10\%$ of the impurities turn out to have nonzero $n_a = \pm 1$. We do not observe $|n_a|$ bigger than 1. Initial states with randomly chosen $\{ n_a \}$ from $\{-1,0,1 \}$ are always trapped in local minima with high energy. Typical phase configurations are depicted in Fig.~\ref{phase}(b1), whose impurity distribution is the same as that of Fig.~\ref{phase}(a1). We can see from this figure that allowing $n_a \neq 0$ removes strong local strains by introducing large fluctuations on a larger length scale than $l$. Thus, the phase fluctuations of size $l$ seen in Fig.~\ref{phase}(a1) disappear, leaving a bigger structure in Fig.~\ref{phase}(b1). The number of vortex-antivortex pairs increases significantly from $n_a = 0$ solutions ($n_V \approx 1.4\%$) [Fig.~\ref{phase}(b2)]. 

\begin{figure}[!tb]
  \begin{center}
   \includegraphics[width = 0.7\columnwidth ]{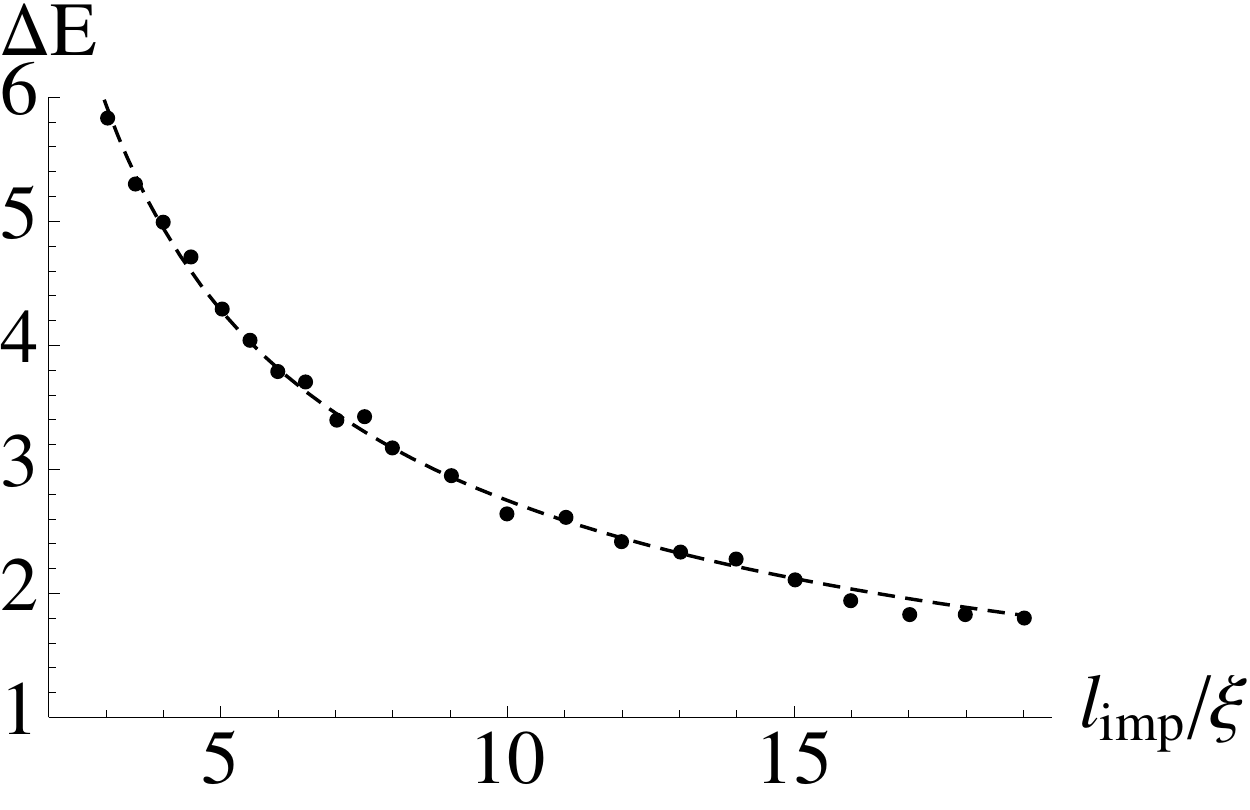}
   \caption{The impurity density dependence of the energy gain for each phase slip $n_a = 0 \rightarrow \pm 1$. The dashed line is the power-law fit given in Eq.~\eqref{fitting} }
    \label{n_var}
  \end{center}
\end{figure}

The average energy gain due to a change of one of the  $n_a$ by unity depends on the concentration of the impurities; it is well fitted by (Fig.~\ref{n_var})
\begin{equation}
\Delta E (l/\xi) \approx \frac{12.0}{(l/\xi)^{0.64}},
\label{fitting}
\end{equation}
in the unit of $2\pi \rho_S \xi$. Considering the number of $n_a \neq 0$, this is a significant change in total energy; the  solution with all $n_a=0$  is of much higher energy than the state we found.  We note that sites with $n_a\neq 0$ occur  mostly when $\theta_a$ is close to $\pm \pi$, while other impurities nearby prefer $\mp \pi$. Since the number of impurities inside the sphere of radius $r_\text{TF}$ grows as $l$ increases, the fluctuation of local phase at the center of the sphere, which induces the change of $n_a$, decreases; thus, the energy gain due to changing the value of  $n_a$ at a single site also decreases as $l$ becomes bigger. 

Although the energy of the $n_a =0$ solution is higher than that of  the true ground state, the conclusion that the system has long-range order does not change even if modifications of $n_a$ are allowed. For example, when we calculate the phase-phase correlation, now the average over the impurity phases is replaced by
\begin{equation}
\langle (\theta_a + 2 \pi n_a) (\theta_b + 2 \pi n_b) \rangle. 
\end{equation}
We find that the autocorrelation of $n_a$'s decays almost immediately over the average inter-impurity distance $l$, indicating $\langle n_a n_b \rangle \propto \delta_{ab}$. Similarly the correlation between $\theta_a$ and $n_a$ is found to be local. Thus, essentially the calculation leading to the long range order does not change. The numerically obtained autocorrelation indeed indicate long ranged order, although the final value is much smaller than the value without $n_a$ variations due to the larger phase fluctuations (Fig.~\ref{auto3}).

\section{Monte Carlo Results \label{MC}}
In order to test our analysis presented so far, and to further understand the physics of the model, we study the phase only model, Eq.~\eqref{phase_model}, placed on a regular periodic lattice by a Monte Carlo (MC) simulation. The model is now mapped to a three dimensional XY model with random and dilute magnetic fields,
\begin{equation}
E = - \sum_{\langle i,j \rangle} \vec{s}_i \cdot \vec{s}_j - \sum_{i \in \{a\}} \vec{h}_i \cdot \vec{s}_i, 
\label{XY_model}
\end{equation}
with adjacent sites $\langle i,j\rangle$, random impurity sites $\{a \}$, and unit vectors $\vec{s}_i$. We consider the $h_i = \infty$ limit with the orientations of $\vec{h}_i$ randomly chosen in the range $[-\pi, \pi]$. This model captures the same large-distance physics as the phase-only model, Eq.~\eqref{phase_model}, while the short-distance properties are different; in particular we expect that $\kappa$ in Eq.~\eqref{kappa} is smaller. 

We perform standard Metroplis Monte Carlo simulations. During the MC steps, we do not update the spins on impurity sites. We will count eight overrelaxation steps and two Metropolis steps as one MC step.\cite{Li1989} Starting from a cold initial condition, where all $\theta_i = 0$, the first $2 \times 10^3$ MC steps are used for thermalization, and the following $1 \times 10^4$ MC steps are used for measurements. We recorded observables every 25 MC steps. We have found that the autocorrelation time is at most 10 MC steps regardless of the size and temperatures; it is relatively short due to the overrelaxation steps. The range of the acceptance rate for Metropolis steps is from 40\% to 60\%.

First, we focus on a low temperature, $T = 0.1$, to compare with the solutions of Eqs.~\eqref{kernelmomentum} and \eqref{J}. Typical phase configurations is given in Fig.~\ref{phase}(c1) given the same impurity configurations as Figs.~\ref{phase}(a1) and (b1). While both vortex-antivortex pairs and $n_a$ variations remove the local strain and lead to a similar structure in phase configurations, a small number of vortex-antivortex pairs in MC ($n_V \approx 0.5\%$) does not induce much phase fluctuations as $n_a$ variations; the larger value of $E_\text{core}$ in the XY model suppresses vortex-antivortex pairs.  In Fig.~\ref{auto3}, the autocorrelations based on Eq.~\eqref{Cx} are plotted for $L= 32$ and $L=64$. Compared to the autocorrelations from the GL analyses, which show rapid saturation to non-zero values, the ones of the MC simulations show a relatively slow decay with no apparent saturation.  Furthermore, while there is very little size dependence in the GL calculations, the MC results depend on the size.

\begin{figure}[!tb]
  \begin{center}
   \includegraphics[width = 0.9\columnwidth]{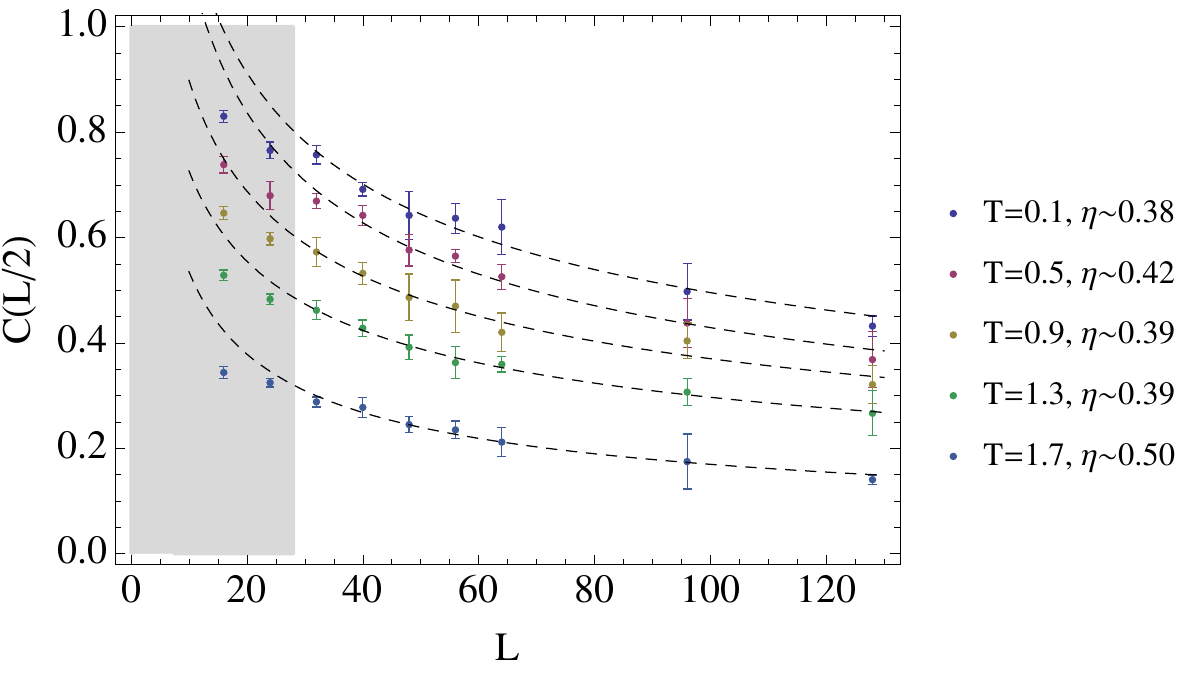}
   \caption{(Color online) Size dependence of autocorrelations obtained by Monte Carlo simulations at $n_\text{imp} = 4\%$ at various temperatures. Only the data for $L \geq 32$ is used for fitting; the data for $L=16$ and $L=24$ (in the gray area) deviates from the power-law behavior in the log-log plot due to their small sizes.}
     \label{auto}
  \end{center}
\end{figure}

To further investigate this size dependence, we look at the end point value of the autocorrelations $C(L/2)$ from $L=16$ to $L=128$ at various temperatures with $n_\text{imp} = 4\%$ (Fig.~\ref{auto}). For temperatures well below $T_c \approx 1.9$, this is well fitted by a power-law as:\cite{Binder2012}
\begin{equation}
C(L/2) \sim L^{-\eta} , \ \ (\eta \approx 0.4) 
\end{equation}
indicating quasi-long-range order. Close to the transition, $T=1.7$, the exponent increases to $\eta  \approx 0.5$. Similar size dependence are checked on magnetization with negligible difference on the values of $\eta(T)$. Therefore, up to the size available for our simulations, our MC results indicate that the low-temperature phase has quasi-long-range order,  with a low density of topological defects, i.e., a Bragg glass.  However the Bragg glass phase found here differs markedly from the conventional Bragg glass phase with weak disorder;\cite{Giamarchi1994, Giamarchi1995} the correlation length exponent we find $\eta \sim 0.4$ is much smaller than the conventional value $\eta_\text{BG} \sim 1.0$. 

We do not have a definitive interpretation of the differences between our MC results and the results obtained by energy minimization, nor a clear account of the difference in exponent relative to the conventional Bragg glass phase. One possibility is that the behavior revealed in the MC calculations is a crossover, visible over a much wider range because the short length scale physics of the XY model is different from that of the phase-only model. In this scenario the anomalously small exponent would be an artifact of the crossover region, with the true behavior being a saturation at longer scales.  Determining the behavior at longer length scales is an  important open question.

\section{Conclusions \label{Conclusions}}
In summary, we investigated the effect of dilute (inter-impurity distance long compared to coherence length) but strong impurities on a CDW state. We find that an effect analogous to screening arising from the long range nature of the elastic forces associated with the pure CDW state substantially suppresses phase fluctuations and leads to a state with long ranged order. The state is characterized by a screening length parametrically larger than the inter-impurity distance. Numerical simulations of two models (the phase-only model, and a XY model), whose small wavelength limits are the same, confirm this basic picture. However, the quantitative behaviours are found to be  different.  For the phase-only model, we find long-range order without size dependence. On the other hand, our Monte Carlo simulations of a XY model indicate that the low temperature state is characterized by power-law correlations with an unusual exponent of $\approx 0.4$. We suggest the apparent exponent is a crossover phenomenon due to the finite size effects, but the issue warrants further research. 

Before we conclude, we briefly touch on several relevant Monte Carlo simulations for the three dimensional XY model with random magnetic fields. Gingras and Huse studied a XY model with random magnetic fields applying on all sites. \cite{Gingras1996} Their findings suggest vanishing of vortices at weak magnetic fields at low temperatures. Fisch considered $q=6$ and $12$ Potts models with dilute, but infinitely strong magnetic fields, and found long-range order at low temperatures and quasi-long-range order at intermediate temperatures for $n_\text{imp} = 6.25\%$.\cite{Fisch1997} These results are qualitatively consistent with our basic finding of long ranged order in this random field model.

This work was supported by Department of Energy Contract Nos. DE-FG02-04ER46157 (J.O.), and DE-FG02-04ER46169 (A.J.M.). We appreciate helpful  discussions with Carlos Arguello, Thierry Giamarchi, Ethan Rosenthal, Susan Coppersmith  and Abhay Pasupathy.

\end{document}